\documentclass[12pt,preprint]{aastex}

\slugcomment{}

\shorttitle{Size of the Narrow Line Region}
\shortauthors{Yonehara}

\begin{document}


\title{Constraining the size of the narrow line region in distant quasars}


\author{Atsunori Yonehara \altaffilmark{1} \altaffilmark{2}}
\affil{Theoretical Astrophysics Group, Department of Physics, 
The University of Tokyo, Hongo 7-3-1, Bunkyo-ku, Tokyo, 113-0033, Japan}
\email{yonehara@utap.phys.s.u-tokyo.ac.jp}


\altaffiltext{1}{Inoue Fellow}
\altaffiltext{2}{Currently, the author affiliates to ARI/ZAH 
(Heidelberg, Germany) by JSPS Postdoctoral Fellowships for Research Abroad. 
E-mail address is {\tt yonehara@ari.uni-heidelberg.de} .}


\begin{abstract}

We propose a proper method to measure 
the size of the narrow line region (NLR) in distant quasars. 
The apparent angular size of the NLR is, in general, 
too small to resolve technically. 
However, it is possible to map the NLR if with gravitational lensing. 
In our method, we directly compare the observed image of the NLR 
with the expected lensed images of the NLR 
for various source sizes and lens models. 
Seeking the best fit image via the comparison procedures,  
we can obtain the best-fit size and the best-fit lens model.  
We apply this method to the two-dimensional spectroscopic data 
of a famous lensed quasar, Q2237+0305.    
If the lens galaxy resembles the applied lens model, 
an upper limit to the NLR size can be set $750~{\rm pc}$.  
Further, we examine how the fitting results will be improved 
by future observations, taking into account 
the realistic observational effects, such as seeing. 
Future observations will provide us more stringent constraints 
on the size of the NLR and on the density profile of the lens galaxy.

\end{abstract}




\section{Introduction}

It is widely believed that Seyfert galaxies and quasars have 
a rather complex structure in their nuclei. 
The central engine seems to be a combination of 
a supermassive black hole and an accretion disk, 
which are surrounded by the broad-line regions (BLRs), 
dusty tori, and the narrow line regions (NLRs) 
\citep[e.g.,][]{ant93}. 
Therefore, investigating the nature of the NLR is useful 
to obtain good insight into the physics of the central engine 
and its environments. 
So far, various kinds of observational programs 
have been carried out to understand the basic physical processes 
involved with the gas in the NLRs, such as photo-ionization, 
shock excitation, and kinematics. 
\citet{sch03}, for example, performed narrow band imaging observations 
of near-by Seyfert galaxies by using 
spatially extended [O~III] $\lambda~5007$ emission line 
and measured the size of the NLRs.  
They found a correlation between the size ($R_{\rm NLR}$) and 
the luminosity ($L_{\rm [O~III]}$) of the NLRs as 
$R_{\rm NLR} \propto L_{\rm [O~III]}^{0.33 \pm 0.04}$,  
which significantly differs from that of quasars \citep{ben02};  
$R_{\rm NLR} \propto L_{\rm [O~III]}^{0.52 \pm 0.06}$.   
On the other hand, \citet{net04} performed 
slit spectroscopy of distant quasars, 
and claimed the existence of two distinct populations 
for luminous active galactic nuclei;  
that is, quasars with and without the NLR. 
We here note, however, that such arguments are purely 
based on the empirical relation 
between $R_{\rm NLR}$ and $L_{\rm [O~III]}$  
obtained from a relatively small number of samples \citep{ben02}. 
It is not yet clear if this relation can be accurately extrapolated 
to luminous active galactic nuclei and/or high-redshift quasars. 
Therefore, direct measurements of the size of the NLRs in luminous 
and/or high-redshift quasars are necessary to confirm their findings 
and to explore the underlying physics. 
However, it is not easy to make direct measurements for 
distance sources, since the more distant the quasars are, 
the smaller the apparent size of the NLRs becomes. 
To make matters worse, it is hard to measure 
the size from narrow band imaging observations,  
because the wavelength of the emission line is redshifted 
and move out from the passband of existing narrow band filters. 

We, here, consider an alternative method which utilizes 
the gravitational lensing, since the gravitational lensing 
can spatially stretch the source image. 
In principle, the size of the NLRs in multiple quasars 
can be measured more precisely, compared with 
that of the unlensed quasars with the same intrinsic luminosity. 
Thus, there have already been such attempts 
by using a two-dimensional spectrograph
attached to a ground based telescope \citep{ada89}. 
  
\citet{med98} performed  two-dimensional spectroscopic observations 
of a lensed quasar with quadruple image, Q2237+0305 (or Huchra's lens),   
by using an optical fiber spectrograph, INTEGRAL. 
They focused on an arc-like feature in the image of 
not [O~III] $\lambda~5007$ but C~III] $\lambda~1909$ emission line 
and derived the size of the NLR in Q2237+0305 to be 
$\sim 400h^{-1}~{\rm pc}$.   
\citet{mot04} used the same observational data, but 
decomposed the spectra into two components;  
the broad line component and the narrow line component. 
They then made two maps for the two spectral components, 
finding a significant difference between them in terms of 
the spatial extent of the line emitting regions. 
After making two images, they focused on the arc-like feature again  
and derive a size of the NLR to be $700 \sim 900~{\rm pc}$.  
This value is consistent with that obtained by \citet{med98}. 
\citet{way05}, in contrast, could not find an arc-like feature 
in the data taken by the GMOS Integral Field Unit (IFU), 
claiming that the size of the NLRs is significantly smaller 
than that obtained by \citet{med98}. 
They remove the seeing effects via the deconvolution technique, 
but did not decompose the spectra into 
the narrow line and the broad line component.
Since their analysis are different not only in the used data 
but also in the adopted analysis procedures, 
we cannot simply compare their results 
nor conclude which is more appropriate. 
Importantly, the lens model degeneracy, 
which was pointed out by \citet{wam94}, 
was not properly considered in these works. 
 
In this paper, we consider a more reliable method to measure 
the size of the NLRs in lensed quasars, 
taking into account the effect of seeing and the lens model degeneracy 
in a proper way and then apply our method to 
both real data and simulated data.  
In the next section, we explain our method 
to measure the size of the NLRs. 
The results obtained by the currently available data 
is presented in section 3. 
In section 4, we examine a potentiality of our method  
for future observational data. 
The final section is devoted for concluding remarks. 
A concordance cosmology with $\Omega_m=0.3$, $\Omega_\Lambda=0.7$, 
and $H_{\rm 0}=70~{\rm km~s^{-1}~Mpc^{-1}}$ (or $h = 0.7$) 
is used throughout this paper.

\section{Our Method}

\subsection{Procedures}

For a given emissivity profile of the source and for a given lens model, 
we can calculate an ideal image of the extended lensed source 
(see next subsection for the lens model).
Convolving the ideal image with a point spread function 
which represents the seeing effect, 
we can obtain an expected image for the observation 
with an infinite spatial resolution. 
Re-sampling this expected image with a finite spatial resolution, 
which is same as the spatial sampling rate of an observational instrument, 
we can produce an image expected by the given observational instrument.
By calculating such images for various source sizes and 
various lens models, and comparing them with the observed one, 
we can finally obtain the best-fit source size and 
the best-fit lens model.  

Here, we apply $\chi ^2-$ minimization method 
to find the best-fit solution. 
The total $\chi ^2-$ is calculated as,   
\begin{equation}
\chi ^2 = \sum_{i=1}^{N_{\rm obs}} \frac{ \left( f_{{\rm obs}, i} 
 - f_{{\rm model}, i} \right) ^2}{\sigma _i^2}, 
\label{eq:chisq}
\end{equation}
where $N_{\rm obs}$, $f_{{\rm obs}, i}$, $f_{{\rm model}, i}$, 
and $\sigma _i$ are the number of data points, 
the observed flux of the $i$-th data point, 
the $i$-th predicted flux from a given model, 
and the observational error of the $i$-th data point, respectively.  

Unless otherwise specified, 
we assume that the source is circularly symmetric  
and that the emissivity profile of the source ($\epsilon$) is 
expressed by Gaussian; 
\begin{equation} 
\epsilon (r) \propto \exp \left( -\frac{r^2}{R_{\rm NLR}^2} \right),  
\label{eq:emis-prof}
\end{equation}
where $r$ represents the distance from the source center. 

Further, we assume that the point spread function 
due to the seeing effect is also circularly symmetric 
and has a Gaussian profile with the same 
``full width half maximum'' as that of the seeing size.

\subsection{Lens model}

It is too hard to measure the size of the NLR 
without any constraint on the lens model, 
we put a practical assumption on the model 
that the mass density profile of the lens galaxy is elliptical 
and is expressed by a power-law in radial profile.
Since there exists the so-called 
lens model degeneracy \citep[e.g.,][]{wam94} even in such situation,  
we need lens models with various density profiles 
to find reliable constraints on the size of the NLR.    
For simplicity, we adopt so-called 
``generalized pseudo-isothermal elliptic potential'' or 
``tilted Plummer family of elliptic potential'' \citep{bla87}. 
In this model, the lens potential ($\phi$) at an image position of 
$\vec{\theta} = (\theta_x, \theta_y)$ is given by  
\begin{equation}
\phi (\vec{\theta}) = \alpha_{\rm E}^{2-2\lambda} \left[ 
\theta_{\rm c}^2 + \left( 1-e \right) \theta_x^2 
+ \left( 1+e \right) \theta_y^2 \right]^{\lambda}, 
\label{eq:lenspot}
\end{equation}
where $\alpha_{\rm E}$, $e$, $\lambda$ and $\theta_{\rm c}$ represents 
the lens size, the ellipticity, the power index and the core radius 
of the lens potential, respectively. 
There are other models that represent ellipticity of the lens galaxies, 
such as softened power-law elliptical mass distribution model 
\citep{bar98}, or other similar lens models with power-law 
density profile \citep{sch90, kee01}. 
Even more sophisticated lens models that consist from bulge, disk, 
and halo component can be used in this kind of studies. 
It is also interesting to perform the same study  
by using different type of lens models, 
but main purpose of the current study is to 
investigate how the method works. 
Moreover, such models require relatively much computational expense, 
and in a current paper, we only focus on a lens model which 
represented by equation~\ref{eq:lenspot}. 
Dependence on different type of lens models 
will be discussed in a future paper.   

A model with large (small) $\lambda$ corresponds to 
a shallow (steep) density profile.  
In particular, the density profile with $\lambda=0.5$ is 
identical to that of the singular isothermal sphere model. 
Since the fifth image has not been detected yet \citep{fal96}, 
we set $\theta_{\rm c}=0$ which is equivalent with 
the lack of the fifth image for this model. 
As was claimed by \citet{kas93}, 
$e > \frac{\lambda}{3-\lambda}$ should be satisfied in this model, 
since otherwise the iso-density contour shows concave shape and 
such model must be unrealistic for a single lens object.  

\begin{figure}
\epsscale{.80}
\plotone{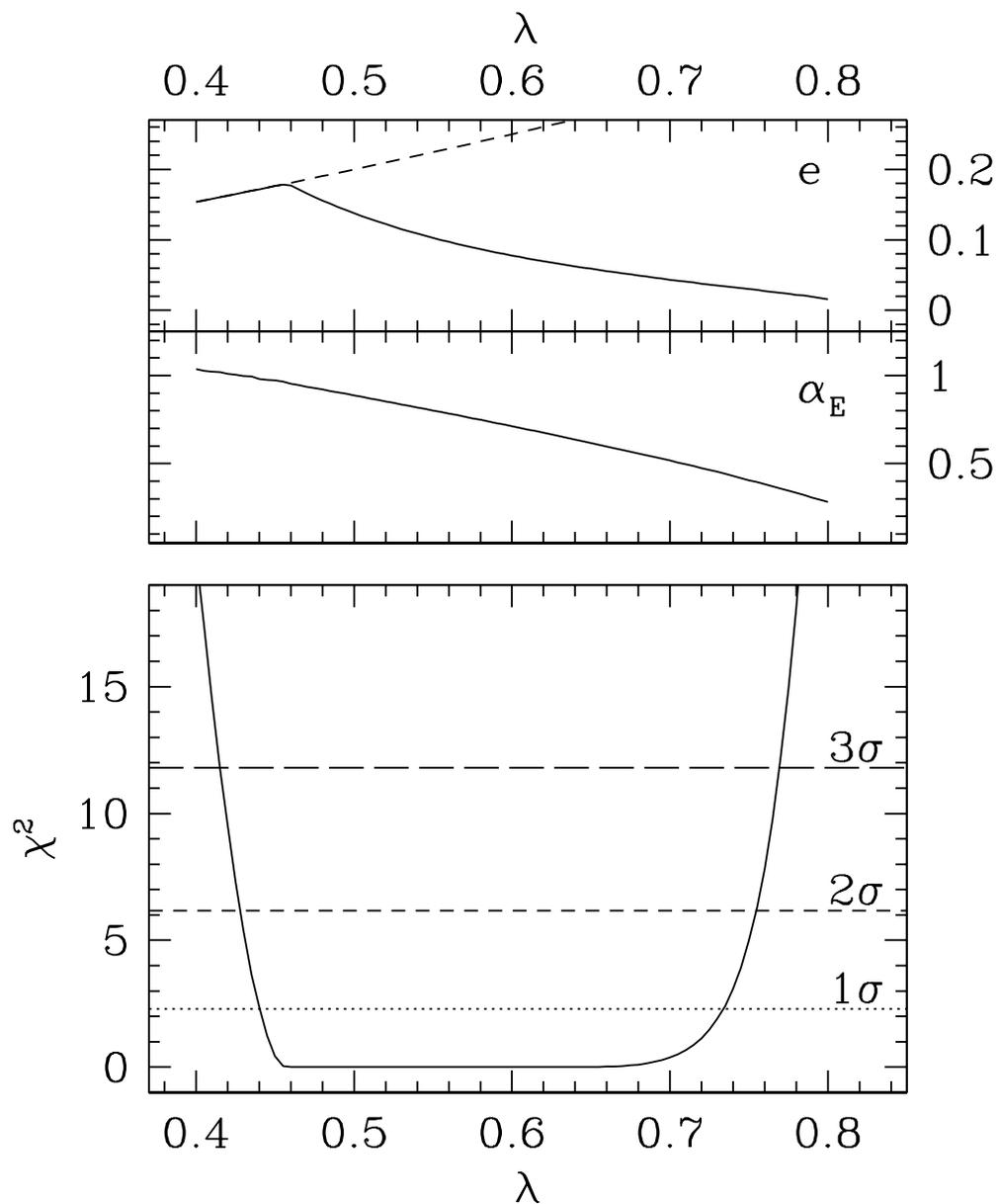}
\caption{Total $\chi ^2$ value (the bottom panel), 
the best fit value of $\alpha _{\rm E}$ 
in the unit of arcsec (the middle panel), 
and that of $e$ (the top panel) is presented with solid line 
as a function $\lambda$.
In the bottom panel, 1-, 2-, and 3-$\sigma$ confidence level is 
also presented by the dotted, dashed, and long-dashed line, respectively.
In the top panel, a critical value of $e$ which is noted by
 \citet{kas93} (see section 2.2) is also presented with the dashed line.}
\label{fig:chi-model}
\end{figure}

For various $\lambda$, we try to reproduce 
the positions of the quadruple image of Q2237+0305  
relative to the position of the lens galaxy 
\footnote{The data are taken from CASTLES website, 
{\tt http://cfa-www.harvard.edu/castles/}.}. 
The number of observable is $2 \times 4 = 8$, 
i.e., $x-$ and $y-$ coordinate of the four images. 
Since observed fluxes may be affected by microlensing and/or 
differential dust extinction, 
we do not take into account the flux ratios between the four images 
in our fitting procedure.  
The number of free parameters is $6$ that is, 
$\alpha_{\rm E}$, $e$, $\lambda$,  
$x-$ and $y-$ coordinate of the source, 
and the direction of the major axis of the lens model.  
Thus, degree of freedom in this fitting procedure is $8 - 6 = 2$. 
$\chi ^2$ values of the best-fit lens model for various $\lambda$ 
are presented in the bottom panel of figure~\ref{fig:chi-model}.  
The resultant values of $\alpha_{\rm E}$ and $e$ which obtained by 
this fitting procedure is also presented in the middle and the top 
panel of figure~\ref{fig:chi-model}, respectively.  
Within 3-$\sigma$ confidence level,  
a wide range of $\lambda$, $\lambda = 0.41 \sim 0.77$, is acceptable 
for the model to reproduce the observed image positions. 
Even if we restrict 1-$\sigma$ confidence level,  
the model with $\lambda = 0.44 \sim 0.73$ provides an acceptable fit, 
and the acceptable range does not shrink so much. 
This represents the lens model degeneracy as was found through 
the previous studies. 

From this fitting procedures, we obtain 
the best-fit parameters of the lens model for a given $\lambda$ 
as shown in the middle and the top panel of figure~\ref{fig:chi-model}. 
Hereafter, we treat the density profile (or $\lambda$) 
as a single free parameter for the lens model in our proposed method; 
once we change $\lambda$ value, other parameters such as $e$ 
are automatically replaced with the best-fit parameters 
for new $\lambda$ value.
We put a constraint on $e$ value to keep a condition 
that is claimed by \citet{kas93} during the fitting procedures, 
i.e., $e \le \frac{\lambda}{3-\lambda}$.
We can see the effect of this constraint 
in the top panel of figure~\ref{fig:chi-model}. 
Basically, the best fit value of $\alpha _{\rm E}$ and $e$ becomes lager, 
when the applied $\lambda$ value for the fitting procedure becomes smaller. 
Though the best fit value of $\alpha _{\rm E}$ gradually increases 
with decreasing $\lambda$ until the smallest value of $\lambda$ 
for the fitting ($\lambda = 0.4$), that of $e$ faces 
the limiting value of the constraint on the fitting, 
i.e.,  $e = \frac{\lambda}{3-\lambda}$, at $\lambda \sim 0.46$. 
Below this $\lambda$ value, the best fit value of $e$ is almost 
identical to the limiting value of $e$.
Therefore, we would like to mention that the validity of 
the fitting results at this small $\lambda$ regime is somewhat uncertain.

\begin{figure}
\epsscale{.80}
\plotone{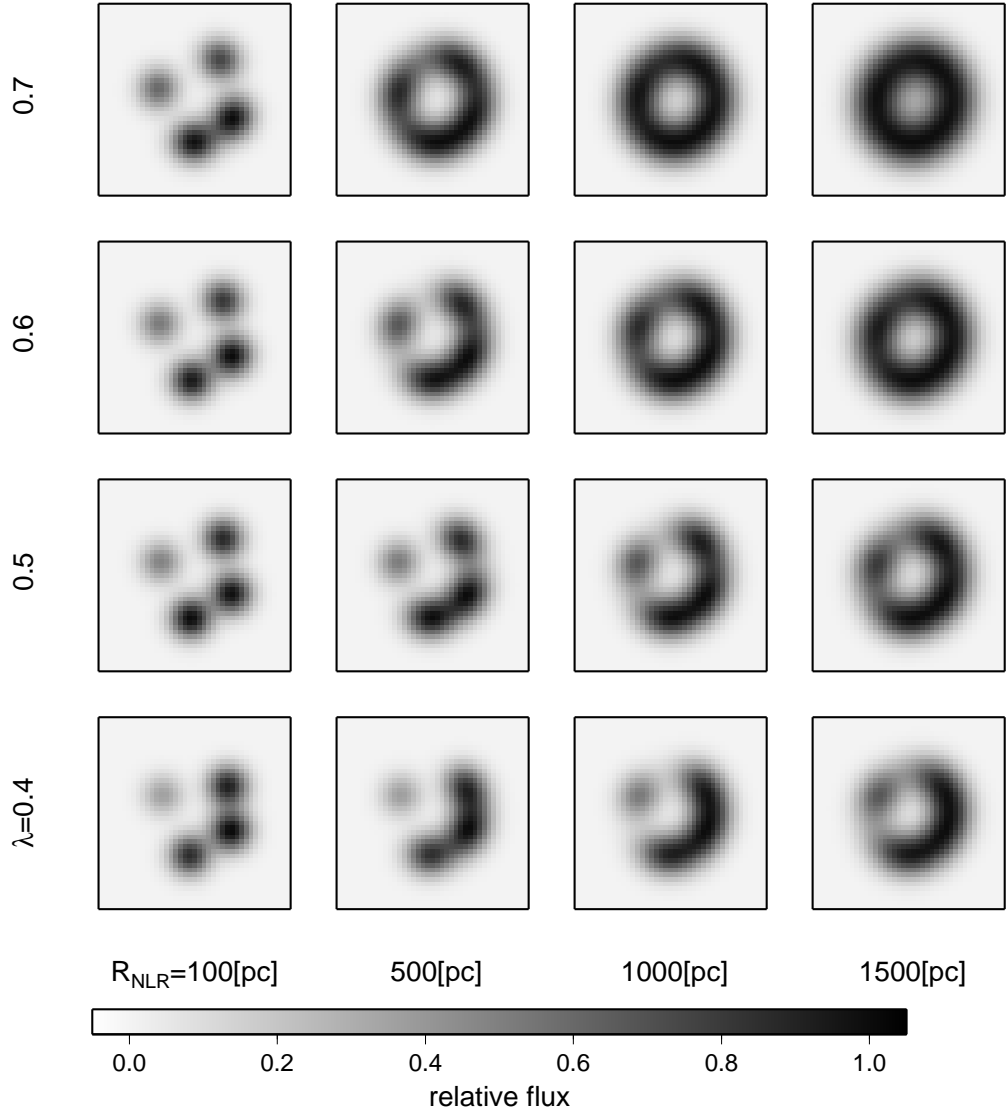}
\caption{Calculated images that are to be 
compared with observational data. 
We indicate the values of $R_{\rm NLR}$ and $\lambda$ 
in the bottom and in the left of the panels, respectively. 
The flux of each pixel is normalized by the peak flux. 
We assume $0.7~{\rm arcsec}$ seeing 
and $0.1~{\rm arcsec}$ sampling rate. }
\label{fig:example}
\end{figure}

Finally, $R_{\rm NLR}$ and $\lambda$ 
is the parameter to be determined from our method.  
Calculated ideal images, which are to be compared with 
the observed image of Q2237+0305, are shown in figure~\ref{fig:example}. 
In the case of large $R_{\rm NLR}$ and/or large $\lambda$, 
the expected image tends to exhibit an arc- or ring-like morphology, 
as is clearly seen in figure~\ref{fig:example}.  
Even if we consider the seeing effect, obviously,  
we can recognize differences between the images for 
the different parameter sets. 
Therefore, we can, in principle, 
constraint the size of the NLRs and the lens model 
via our proposed method.

\section{Analyses using the Current Data}

\citet{mot04} have performed Gaussian fit to 
the emission line profiles in observed spectra, 
and have decomposed the detected emission line into 
a broad line component and a narrow line component.  
The resultant line intensity peak ($I$) and line width ($w$) of 
both components of the C~III] $\lambda~1909$ emission line  
are shown in table 1 of \citet{mot04}.
Since total flux of an emission line with Gaussian profile 
is calculate by $\sqrt{\pi} I w$, 
we evaluate the total flux ($\sqrt{\pi} I w$) of 
both emission line components at each fiber 
from the quantities, $I$ and $w$, presented in table 1 of \citet{mot04}.
Error of flux at each fiber is estimated by using 
the error of the line intensity peak($\delta I$) and that of  
the line width ($\delta w$) as usual manner; the error is equal to 
$\sqrt{\pi} \sqrt{I^2 \cdot \delta w^2 + w^2 \cdot \delta I^2}$. 
Consequently, we can estimate the goodness of fit, 
i.e., the $\chi ^2$ value, between a model and 
the observational data via equation~\ref{eq:chisq}. 
This procedure could be applicable if the emission line was detected, 
and if the decomposition into the broad line component 
and the narrow line component was succeeded.  
However, the emission line was not detected in some fibers, 
e.g., fiber 107 of \citet{mot04}. 
In such case, the total observed flux is set to be zero and 
the error is set to be the same as the error of a fiber 
that the total flux is the weakest in emission line detected fibers. 
The $\chi ^2$ value of such fibers can also be estimated 
via equation~\ref{eq:chisq}. 
Additionally, even if an emission line is detected, 
decomposition into the broad line component and the narrow line component 
might not be succeeded in some fibers, e.g., fiber 105 of \citet{mot04}.     
In such case, the emission line is treated as a single component 
in table 1 of \citet{mot04}, and the total observed flux 
is set to be an upper limit for both components and  
the error for both components is evaluated from the same observational data. 
If a model flux of a fiber exceeds the observed flux of the fiber, 
the $\chi ^2$ value of the fiber is evaluated by equation~\ref{eq:chisq}. 
If a model flux of a fiber is below the observed flux of the fiber, 
the $\chi ^2$ value of the fiber is set to be zero. 

The spatial sampling rate is set to be the same as INTEGRAL, and 
it is roughly $0.5~{\rm arcsec}$ in the case of \citet{mot04}. 
Both of the shape of a fiber cross section and the flux loss between fibers 
were taken into account. 
As reported in \citet{mot04}, seeing is set to be $0.7~{\rm arcsec}$. 
We make use of the data of $6 \times 6 = 36$ fibers, in total, 
in a rectangular array. 
Since the number of the fitting parameter is two; 
$R_{\rm NLR}$ (or $R_{\rm BLR}$ for BLR) and $\lambda$, 
degree of freedom in the fitting is $36 - 2 = 34$. 
The following results are summarized 
in the upper part of table~\ref{tab:summary}.

\subsection{Broad line region}

The Gaussian intensity peak for the broad line component 
is presented in table 1 of \citet{mot04}, 
but the Gaussian width for the broad line component is not. 
Thus, following the analysis by \citet{mot04}, 
we set the width to be $71 \pm 11~{\rm \AA}$ for all fiber. 
The searched parameter ranges of $R_{\rm BLR}$ and $\lambda$ are  
$50 \sim 2050~{\rm pc}$ and $0.4 \sim 0.8$, respectively. 
The best fit parameter search is performed 
with $50~{\rm pc}$ resolution in $R_{\rm BLR}$ 
and $0.01$ resolution in $\lambda$. 
The result is presented in figure~\ref{fig:motta}a. 
The best fit parameters are $R_{\rm BLR} = 50~{\rm pc}$ 
and $\lambda = 0.47$ ($\chi ^2 = 28.01$). 
However, even if we allow 1-$\sigma$ confidence level, 
a wide range of $\lambda$ values are acceptable  
and $R_{\rm BLR}$ cannot be constrained tightly. 
To get more reliable result, 
we had better to choose 3-$\sigma$ confidence region,  
rather than 1-$\sigma$ confidence region. 
Unfortunately, in such case, most of the parameter set 
is acceptable as seen in figure~\ref{fig:motta}a, 
and we can say nothing about the lens model and 
the size of the BLR. 
Consequently, we cannot put any useful constraint 
on density profile of the lens model from the available data.  
We can only put an upper limit to the size of the BLR  
as $R_{\rm BLR} < 400~{\rm pc}$ in 1-$\sigma$ confidence level.

\subsection{Narrow line region}

\begin{figure}
\epsscale{.80}
\plotone{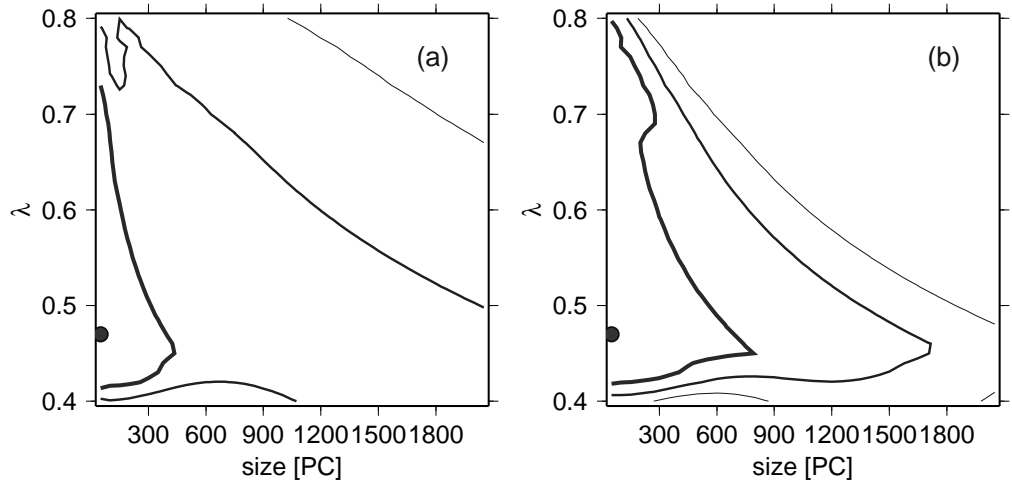}
\caption{The results of fitting to the broad line component (a: left panel)
and the narrow line component (b: right panel), respectively.
The thick, middle, and thin lines present 1-, 2-, and 3-$\sigma$ 
confidence levels, respectively. 
The filled circle indicates the best fit parameter.}
\label{fig:motta}
\end{figure}

We use the same technique to investigate NLRs. 
The Gaussian intensity peak for the narrow line component 
and the Gaussian width for the narrow line component 
are taken from table 1 of \citet{mot04}. 
The searched parameter ranges and the resolution 
are the same as those in the case for the BLR. 
The fitting result is presented in figure~\ref{fig:motta}b. 
Again, the best fit parameter is $R_{\rm NLR} = 50~{\rm pc}$ 
and $\lambda = 0.47$ ($\chi ^2 = 25.09$), but 
wider ranges of $R_{\rm NLR}$ and $\lambda$ are acceptable. 
However, there are two major differences 
between figure~\ref{fig:motta}a and b. 
Firstly, the upper limit on $R_{\rm NLR}$ is 
significantly larger than that on $R_{\rm BLR}$  
in 1-$\sigma$ confidence level. 
Secondly, the parameter ranges in the upper right part 
of figure~\ref{fig:motta}b must be rejected, 
even if we allow 3-$\sigma$ confidence level. 
Unlike \citet{mot04}, we are not able to 
measure $R_{\rm NLR}$ from the data. 
Its upper limit is $\sim 750~{\rm pc}$ in 1-$\sigma$ confidence level. 
Again, we are not able to put useful constraints on 
the lens model as in the case of the BLRs.

\section{Analyses by using Improved Data}

From the fitting results presented in the previous section, 
we understand that the currently available data is 
not good enough to measure the size of the NLR 
and to constrain on the lens model. 
One reason for this can be attributed to 
the large spatial sampling rate in the used data, 
but this can easily be improved.  
Rather, better smaller spatial sampling rates 
with the same signal-to-noise (hereafter, S/N) ratio 
have already been achieved technically \citep{way05}. 
For instance, actually, GMOS-IFU on the Gemini North, 
SINFONI on VLT, or KYOTO-3DII on SUBARU telescope \citep{sug04} 
can provide us with two-dimensional spectra 
with $0.1~{\rm arcsec}$ spatial sampling rate. 
Thus, it is worth of testing the potentiality of the proposed method 
for the data with $0.1~{\rm arcsec}$ spatial sampling rate. 
In this section, we make mock observational data, 
which have better sampling rate, for a given parameter set
($\lambda$ and $R_{\rm NLR}$ or $R_{\rm BLR}$), 
and apply our method to those data to see how nicely 
it can reproduce the given parameter values.

\subsection{Our procedures}

Throughout this section, we adopt 
$R_{\rm NLR} = 1000~{\rm pc}$ and $\lambda = 0.5$. 
In this lens model, the best fit value of $\alpha _{\rm E}$ 
and $e$ is $0.887~{\rm arcsec}$ and $0.138$, respectively. 
Then, we can calculate the expected observational images 
for a given spatial sampling rate, 
which is currently set to be $0.1~{\rm arcsec}$. 
Moreover, observational noise is artificially added 
to the expected image, and we obtain mock observational data.  
There may be several sources of noise, 
but we assume so-called photon noise to be dominant. 
Unless otherwise specified, seeing size is fixed to be 
$0.7~{\rm arcsec}$ 
\footnote{This value is somewhat worse compared with 
typical seeing at Mauna Kea site.} 
and the noise at the peak flux is set to be $10~\%$. 
Data from $41 \times 41 = 1681$ fibers are used for fitting. 
Roughly $4~{\rm arcsec} \times 4~{\rm arcsec}$ field is 
covered, which is sufficient for our purpose. 
Since the number of parameters are $2$ ($R_{\rm NLR}$ and $\lambda$), 
degree of freedom for the fitting is $1681 - 2 = 1679$. 
The following results are summarized 
in the lower part of table~\ref{tab:summary}.

\subsection{Seeing effects}

The result is shown in figure~\ref{fig:kyoto3d2}a.
The best fit parameter is $R_{\rm NLR} = 1000~{\rm pc}$ and 
$\lambda = 0.5$ ($\chi ^2 = 1637.47$). 
In 1-$\sigma$ confidence level, 
the acceptable range of the parameters are still large, 
and the lens model degeneracy still remains in part. 
Even so, however, we can put tighter constraints on the parameters 
than for the currently available data (see figure~\ref{fig:motta}). 
The acceptable ranges in 1-$\sigma$ confidence level are  
$R_{\rm NLR} \simeq 600 \sim 1300 {\rm pc}$ and 
$\lambda \simeq 0.45 \sim 0.59$. 
Therefore, we can nicely reproduce the given parameters 
with $\sim 40 \%$ accuracy for $R_{\rm NLR}$ and 
$\sim 18 \%$ accuracy for $\lambda$. 

The result in the case of  
smaller seeing size is shown in figure~\ref{fig:kyoto3d2}b. 
Again, the best fit parameter is $R_{\rm NLR} = 1000~{\rm pc}$ and 
$\lambda = 0.5$ ($\chi ^2 = 1634.92$). 
Areas of all of the three confidence regions are smaller 
than those in the previous case (see figure~\ref{fig:kyoto3d2}a), 
although the differences are small. 
The acceptable ranges in 1-$\sigma$ confidence level are 
$R_{\rm NLR} \simeq 650 \sim 1300 {\rm pc}$ ($\sim 35 \%$ accuracy) 
and $\lambda \simeq 0.45 \sim 0.58$ ($\sim 16 \%$ accuracy). 
Although certain observations with better seeing will provide us 
with somewhat better information, the result do not alter much, 
compared with cases of the data 
with the same S/N ratio but with worse seeing. 

In contrast, if the incorrect seeing size is used in the analysis, 
the situation will become worse. 
The best-fit size is expected to be smaller (larger) than true value 
when we overestimate (underestimate) the seeing size.  
For instance, if we assume $0.6~{\rm arcsec}$ ($0.8~{\rm arcsec}$) seeing 
and perform fitting to the mock observational data 
with $0.7~{\rm arcsec}$ seeing, 
the minimum $\chi ^2$ value will exceed 3-$\sigma$ level, 
and the best-fit size and $\lambda$ will be  
$1500~{\rm pc}$ ($550~{\rm pc}$) and $0.46$ ($0.58$), respectively. 
Thus, it must be crucial to know the seeing effect correctly 
for the success of our proposed method.  
The best-fit values are shifted along the sequence of degeneracy 
which has already been appeared in figure~\ref{fig:kyoto3d2}a. 
Since $0.1~{\rm arcsec}$ corresponds to $850~{\rm pc}$ at 
the redshift of this quasar, 
the difference between the true size and the best-fit size, 
$\sim 500~{\rm pc}$, is somewhat smaller than the difference between 
the assumed seeing size and the true seeing size. 
This may be due to the most essential effect of gravitational lensing, 
image stretching of the observed images, in part.

\subsection{Observational errors}

Next, we examine the effects of observational errors 
on the results by decreasing noise, or increasing S/N ratio, 
at the peak flux to $5 \%$. 
In order to reduce observational errors to this value, 
roughly $\left( 0.1 / 0.05 \right)^2 = 4$ times longer exposure time 
is needed than that in the previous case. 
The result is shown in figure~\ref{fig:kyoto3d2}c. 
Again, the best fit parameter is $R_{\rm NLR} = 1000~{\rm pc}$ and 
$\lambda = 0.5$ ($\chi ^2 = 1648.09$), but 
each the confidence region becomes narrower dramatically. 
The acceptable parameter range in 1-$\sigma$ confidence level is 
$R_{\rm NLR} \simeq 850 \sim 1150 {\rm pc}$ ($\sim 15 \%$ accuracy) 
and $\lambda \simeq 0.48 \sim 0.53$ ($\sim 6 \%$ accuracy).
Even in 3-$\sigma$ confidence level for this high S/N ratio case
that is shown in figure~\ref{fig:kyoto3d2}c, 
the acceptable parameter ranges are somewhat smaller than 
those in 1-$\sigma$ confidence level for lower S/N ratio case 
that is shown in figure~\ref{fig:kyoto3d2}a.

\subsection{Clumpy nature of the narrow line region}

In above studies, we assumed Gaussian emissivity profile 
for the NLRs as given by equation~\ref{eq:emis-prof}. 
In actual Seyfert galaxies, however, the emissivity profile may not be  
smooth but can be very complex; likely to be 
composed of small numerous clumps. 
Here, we assume that the NLRs are composed of $100$ clumps,  
the emissivity profile of each clump obeys Gaussian profile 
(equation~\ref{eq:emis-prof}) with a size of $R=50~{\rm pc}$. 
We assume that the clumps are randomly distributed 
and the distribution of distance between 
the source center and each clump obeys Gaussian distribution 
with the typical size of $1000~{\rm pc}$.

The results are shown in figure~\ref{fig:kyoto3d2}d.
We find that the best fit values of the parameters are 
$R_{\rm NLR} = 1000~{\rm pc}$ and $\lambda = 0.49$ ($\chi ^2 = 1660.11$). 
The best fit value of $\lambda$ is slightly smaller than the given
value, but that of $R_{\rm NLR}$ is identical to the given value.  
The shape of 1-$\sigma$ confidence region is different from 
that in figure~\ref{fig:kyoto3d2}a, but the overall properties,  
such as the direction of the major axis of 1-$\sigma$ confidence region,  
are similar to each other. 
Although the extent of 1-$\sigma$ confidence region in this figure 
is smaller than that in figure~\ref{fig:kyoto3d2}a, 
the given values are included in 1-$\sigma$ confidence region, 
$R_{\rm NLR} = 650 \ sim 1200~{\rm pc}$ and $\lambda = 0.45 \sim 0.56$. 

Therefore, we can safely conclude that 
we can put a reliable constraint on the size of the NLRs 
and the lens model by the proposed method, 
even if the clumpy nature of the NLRs are taken into account.  
The clumpy nature may work as an origin of the systematic error 
in the proposed method.

\section{Concluding Remarks}

In this paper, we present the potentially useful method to constrain 
the size of the NLR and the lens model. 
Taking the case of Q2237+0305 as an example, 
we re-analyze the two dimensional spectroscopic data taken by \citet{med98} 
and obtain reliable constraints on the size of the BLR and the NLR. 
Although both of the sizes are consistent with point-like source, 
the possibility of the extended source is still remains.   
The obtained upper limit to the sizes of the BLR and the NLR are 
$400~{\rm pc}$ and $750~{\rm pc}$, respectively. 
As far as the mass profile of the lens galaxy is nicely represented 
by a model where the lens potential is elliptical and a power-law 
in radial profile, as we assumed in this paper, 
these values can be appropriate upper limits. 

In addition, we have demonstrated that 
the currently available instruments can provide 
a better opportunity to explore the NLRs and the lens model. 
Seeing is an important factor in ground-based observations.  
However, it may not seriously affect our results, 
if we know the seeing effect correctly.  
Observations with higher S/N ratios and  
with higher spatial sampling rates will be  
useful to give better constraints on the parameters.  
Even if the spatial distribution of the NLR is not smooth but clumpy, 
we will be able to reproduce the values of the basic parameters successfully. 

One thing that we have to mention is so-called ``mass sheet degeneracy'' 
in gravitational lensing phenomena; if there is a mass sheet 
with the surface mass density of $\kappa_{\rm s}$ between 
observer and the source, any observed structure will be 
scaled up by a factor of $\left( 1 - \kappa_{\rm s} \right)^{-1}$ in length 
and by a factor of $\left( 1 - \kappa_{\rm s} \right)^{-2}$ in area. 
Consequently, the actual source size should be 
the produce of $\left( 1 - \kappa \right)$ and the obtained source size
\footnote{The resultant density profile does not change due to 
this degeneracy, but the lens size term in equation~\ref{eq:lenspot}, 
$\alpha_{E}^{2-2\lambda}$, should be multiplied by 
$ \left( 1 - \kappa_{\rm s} \right) $ to obtain the actual value.}. 
Though the measurement of $\kappa_{\rm s}$ is practically hard 
and annoying issue, such difficulty may be solved at least in part 
by observational studies of the environments around lens systems 
\citep[e.g.,][]{fau04}. 

\begin{figure}
\epsscale{.80}
\plotone{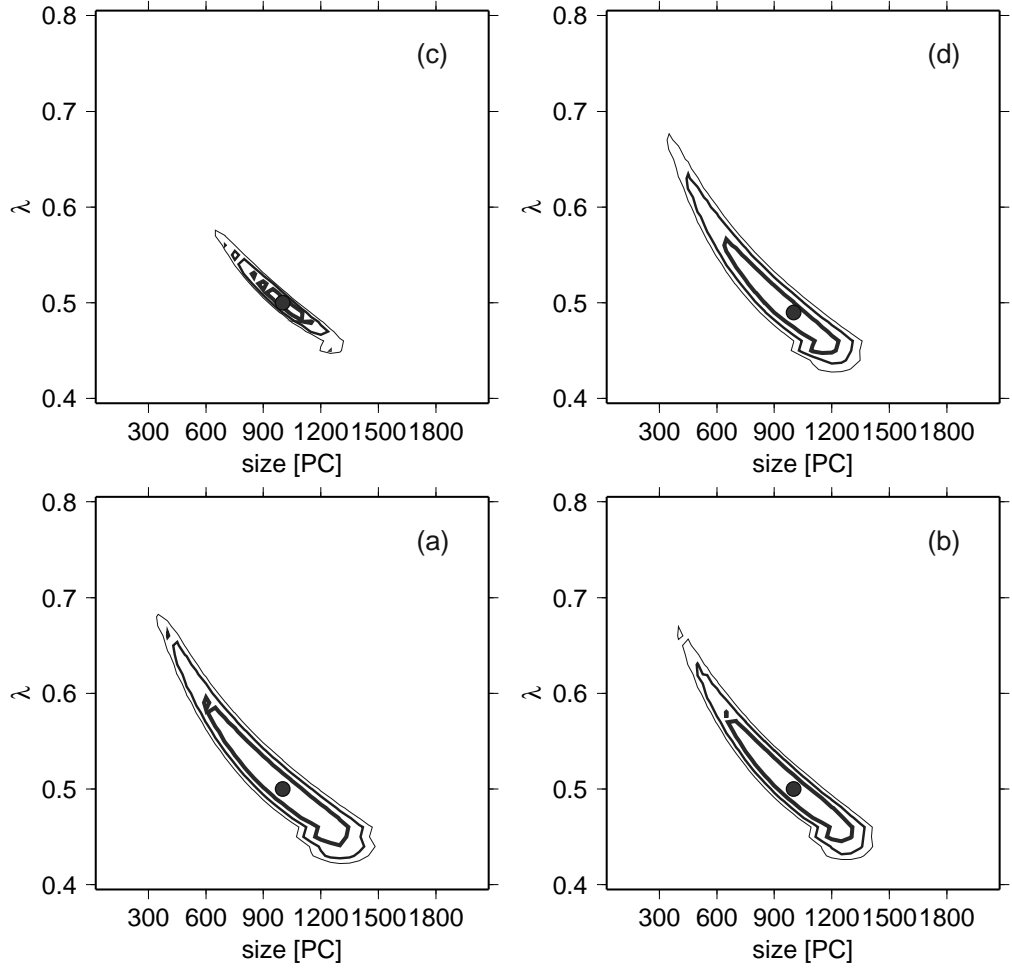}
\caption{Same as figure~\ref{fig:motta}, 
but for the mock observational data (see section 4).
Panel (b) and (c) shows the result for better seeing case and 
smaller error case compared with panel (a), respectively. 
Panel (d) shows the result for the clumpy NLR.}
\label{fig:kyoto3d2}
\end{figure}

If the emissivity profile of the NLR has an elliptical shape 
or a double-cone shape, the lensed image of the NLR can be different 
from those considered in this paper and 
these different properties will alter the results to some extent. 
Further, the existence of substructures in or around 
the lens galaxy may also affect the observed image of the NLR 
\citep[e.g.,][]{met04}.
To evaluate such effects, quantitative estimations are necessary. 
Even in such case, however, we can, in principle, find 
correct values of the parameters in a statistical fashion,  
e.g., by searching the deepest valley of $\chi ^2$ 
(see figure~\ref{fig:kyoto3d2}), 
and to obtain an observational insight into 
the shape of the NLR and/or substructures 
as well as the size of the NLR and the lens model.

~


The author acknowledges to S. Mineshige, H. Sugai, T. Nagao, K. Yahata, 
W.-H. Bian, and anonymous referee for their valuable comments and discussions.


\clearpage

%

\begin{deluxetable}{crrrrc}

\tablecaption{Summary of the fitting results \label{tab:summary}}

\tablewidth{14cm}

\tablehead{ \colhead{data \phd} & \colhead{size\tablenotemark{a}} 
 & \colhead{(1-$\sigma$ range)} & \colhead{\phd $\lambda$} 
 & \colhead{(1-$\sigma$ range)} & \colhead{\phd figure} }

\startdata
BLR\tablenotemark{b} \phd & $50$ & ($50 \sim 400$) & 
 \phd $0.47$ & ($0.42 \sim 0.71$) & \phd figure~\ref{fig:motta}a \\
NLR\tablenotemark{b} \phd & $50$ & ($50 \sim 750$) & 
 \phd $0.47$ & ($0.42 \sim 0.79$) & \phd figure~\ref{fig:motta}b \\
standard\tablenotemark{c} \phd & $1000$ & ($600 \sim 1300$) & 
 \phd $0.50$ & ($0.45 \sim 0.59$) & \phd figure~\ref{fig:kyoto3d2}a \\
better seeing\tablenotemark{c} \phd & $1000$ & ($650 \sim 1300$) & 
 \phd $0.50$ & ($0.45 \sim 0.58$) & \phd figure~\ref{fig:kyoto3d2}b \\
smaller errors\tablenotemark{c} \phd & $1000$ & ($850 \sim 1150$) &  
 \phd $0.50$ & ($0.48 \sim 0.53$) & \phd figure~\ref{fig:kyoto3d2}c \\
clumpy nature\tablenotemark{c} \phd & $1000$ & ($650 \sim 1200$) & 
 \phd $0.49$ & ($0.45 \sim 0.56$) & \phd figure~\ref{fig:kyoto3d2}d \\
\enddata

\tablenotetext{a}{$R_{\rm BLR}$ for the BLR, 
and $R_{\rm NLR}$ for the NLR (unit: pc).}
\tablenotetext{b}{Real observational data obtained by \citet{mot04} 
 (see section 3).}
\tablenotetext{c}{Mock observational data calculated in this paper 
 (see section 4).}

\end{deluxetable}

\clearpage

\clearpage

\clearpage

\clearpage

\end{document}